%% file: main.tex
\documentclass[sigconf]{acmart}

\AtBeginDocument{%
  }

\usepackage{tabularx}

\copyrightyear{2026}
\acmYear{2026}
\setcopyright{cc}
\setcctype{by}
\acmConference[CIKM '26]{Proceedings of the 35th ACM International Conference on Information and Knowledge Management}{November 07--11, 2026}{Rome, Italy}
\acmBooktitle{Proceedings of the 35th ACM International Conference on Information and Knowledge Management (CIKM '26), November 07--11, 2026, Rome, Italy}
\acmDOI{10.1145/3799682.3839977}
\acmISBN{979-8-4007-2539-5/2026/11}

\begin{document}

\title[PairWatch: Pair-Aware Anomaly Transformer for Satellites]{PairWatch: Pair-Aware Anomaly Transformer for Detecting Adversarial Rendezvous in Public Satellite Catalogs}

\author{Young Eun Kwon}
\email{equation@korea.ac.kr}
\orcid{0009-0009-2808-0763}
\affiliation{%
  \institution{Korea University}
  \city{Seoul}
  \country{Republic of Korea}
}

\author{Ji Won Yoon}
\email{jiwon_yoon@korea.ac.kr}
\affiliation{%
  \institution{Korea University}
  \city{Seoul}
  \country{Republic of Korea}}

\renewcommand{\shortauthors}{Young Eun Kwon \& Ji Won Yoon}

\begin{abstract}
\input{sections/00_abstract}

\end{abstract}



\maketitle

\section{Introduction}
\input{sections/01_intro}

\section{Methodology}
\input{sections/02_method}

\section{Evaluation}
\input{sections/03_eval}

\section{Conclusion}
\input{sections/04_conclusion}

\begin{acks}
This work was supported by Korea Research Institute for defense Technology planning and advancement(KRIT) grant funded by the Korea government(Defense Acquisition Program Administration), (KRIT-CT-24-001, Defense Space Security Research Lab, 2024).
\end{acks}

\bibliographystyle{ACM-Reference-Format}
\bibliography{ref}

\end{document}

%% file: sections/00_abstract.tex
Documented adversarial rendezvous and proximity operations (RPO) appear in the public TLE catalog, yet existing TLE-based maneuver detectors treat each satellite in isolation---mistaking routine station-keeping for adversarial behavior and missing RPO entirely when the target stays passive. We argue that the right unit is the satellite pair and the right inductive bias is the kinematic symmetry between approach and recede. We instantiate both with PairWatch, a pair-aware AnomalyTransformer whose attention prior is augmented with a sign-symmetric geometric similarity over relative-motion features, and release an open benchmark of 359 windows from public Space-Track data spanning 11 documented RPO events, 100 Clohessy--Wiltshire synthetic adversarial windows, and a five-class hard-negative set culminating in the TerraSAR-X / TanDEM-X helix formation. Under the deployment-relevant criterion of a single threshold separating real RPO from operational negatives---a necessary property for catalog-scale screening---PairWatch is the only model that succeeds, with the property holding in 100\% of bootstrap resamples versus 0--19.5\% for three neural baselines and under 0.1\% for four physics-informed detectors. A per-source analysis exposes a shortcut-learning failure mode of single-number AUROC: simpler baselines attain higher overall AUROC by shortcutting a synthetic-vs-formation contrast with no operational meaning; removing this single cell flips the test ranking entirely (LSTM-AD), and these baselines fail the security-relevant task the benchmark was built to measure.

%% file: sections/01_intro.tex
Public Two-Line Element (TLE) catalogs maintained by the US 18th Space Defense Squadron are the only globally observable record of the motion of every active satellite, and one of the few remaining open primary sources of space situational awareness. Mining these streams for security-relevant events---not collision risk alone---has gained urgency as documented adversarial proximity operations enter the open record \cite{pavur,falco}: Russia's Cosmos~2542/2543 shadowed the US national-security payload USA~245 in 2019--20, with analogous inspector triplets and the more recent Cosmos~2558 / USA~326 pairing following similar patterns \cite{swf2022,csis2025}. These events are not anomalies to be discovered after the fact---they are an existing operational reality that public catalogs already record, if a detector knows where to look.

Existing TLE-based detection looks in the wrong place. The astrodynamics literature treats each satellite as an independent time series \cite{lemmens} and reduces detection to a single binary outcome---maneuver or no maneuver---on that satellite alone. For rendezvous and proximity operations (RPO), this framing fails by construction. The defining feature is the relative motion between an active attacker and an often-passive target, so a routine station-keeping burn on the target is, in single-satellite view, indistinguishable from no event. A single-satellite Isolation Forest on mean-element residuals \cite{iforest} confirms this, scoring documented real RPO at AUROC~0.228---well below chance. Generic multivariate anomaly detectors applied to pairwise features \cite{omnianomaly,usad} do better but on the wrong subset, distinguishing synthetic RPO from synthetic hard negatives---a contrast with no operational meaning---while consistently confusing real RPO with routine station-keeping and constellation co-orbiting. The right unit of analysis is the satellite pair, and the right inductive bias is the kinematic symmetry that defines an RPO trajectory: a deliberate approach followed by a deliberate recede, linked by a closest approach where relative velocity is at its characteristic minimum.

We address both points with three contributions. (1) PairWatch, a pair-aware AnomalyTransformer whose temporal prior is augmented with a sign-symmetric geometric similarity over pair-kinematic features; the sign-symmetric design halves the across-seed standard deviation of the naive formulation (0.019 to 0.009) without performance cost. (2) An open RPO benchmark: 359 windows from public Space-Track TLE combining 11 documented real adversarial events, 100 Clohessy--Wiltshire synthetic adversarial windows, and a five-class hard-negative set culminating in the TerraSAR-X / TanDEM-X helix formation (120-meter closest approach). (3) A per-source analysis revealing a shortcut-learning \cite{shortcut} failure mode of single-number AUROC: PairWatch is the only model admitting a single threshold separating real RPO from operational negatives, while simpler baselines attain higher overall AUROC by shortcutting a synthetic-vs-formation contrast with no operational meaning---a failure exposed only by per-source breakdown.

%% file: sections/02_method.tex
\subsection{Threat Model}

We consider an adversary that operates a maneuverable satellite $A$ to achieve sustained close proximity with a target satellite $T$ that it does not own, for purposes of signals intelligence or political signaling. The defender observes only public TLE streams; the adversary cannot suppress independent radar tracking that produces these elements.

Given the TLE histories $\mathcal{T}_A$ and $\mathcal{T}_T$ over a fixed analysis window $W$, the detector outputs a score $s(A, T, W) \in [0, 1]$ that estimates the probability that $A$ is conducting adversarial rendezvous and proximity operations toward $T$ within $W$. Catalog-scale scoring follows the standard top-$k$ reduction $s(T, W) = \max_{A \in \mathrm{topk}(T)} s(A, T, W)$, but in this paper we focus on the single-pair task.

\subsection{Pair Feature Vector}

For each pair $(A, T)$ we propagate both satellites with SGP4 \cite{sgp4} at hourly resolution over a 14-day window ($\mathrm{SEQ\_LEN}=336$), and at each timestep $t$ extract a four-channel pair feature vector
\begin{align*}
g_t = \big[\, r(t),\;\dot{r}(t),\;\lVert \vec{v}_A - \vec{v}_T \rVert,\;\theta(t)\,\big]^{\!\top}
\end{align*}
of inertial range, range rate, relative speed, and orbital-plane alignment angle.

We report all numbers in a rate-only setting in which absolute range $r$ is withheld, leaving the three rate channels. Two reasons motivate this. First, the benchmark itself defeats range as a shortcut: the median real-RPO range in our positives exceeds the TanDEM-X closest approach and is comparable to the GRACE-FO geodesy pair \cite{gracefo}. Second, an adversary controls its standoff distance and can dwell at constellation-like separations before closing, so any range-thresholded detector either misses the dwell phase or fires on every close-flying constellation pair. Rate-based detection captures adversarial intent in the kinematic transition, the operationally robust regime our prior targets. Features are z-scored per channel using training-split statistics; SGP4 propagation at uniform hourly timestamps gives consistent A--T epoch alignment without imputation or outlier removal, the 14-day window being short enough that propagation error remains bounded.

\subsection{PairWatch Architecture}

PairWatch builds on the AnomalyTransformer encoder, which maintains at each token both a series association $A_{t,t'}$ (learned self-attention) and a prior association $P_{t,t'}$, and uses the minimax discrepancy between the two as an anomaly criterion at inference, weighted by the per-token reconstruction error~\cite{anomalytransformer}. Vanilla AnomalyTransformer fixes the prior to a purely temporal Gaussian, $P^{(\mathrm{vanilla})}_{t,t'} \propto \exp(-(t-t')^2 / \sigma_t^2)$, which ignores the kinematic structure that defines RPO. We augment it with a geometric term,
\begin{align}
P_{t,t'} \;\propto\; \exp\!\Big(\,-\tfrac{(t-t')^2}{\sigma_t^2}\;-\;\alpha\,\tfrac{\lVert \tilde{g}_t - \tilde{g}_{t'}\rVert^2}{\beta_t^2}\,\Big),
\label{eq:prior}
\end{align}
where $\sigma_t$ is the temporal bandwidth inherited from vanilla AnomalyTransformer, $\beta_t = \mathrm{softplus}(w_\beta^{\!\top}\tilde{g}_t)$ is a per-token geometric bandwidth from a learned projection $w_\beta \in \mathbb{R}^{2F}$, and $\alpha \in [0,1]$ is a learned per-head scalar gate. The geometric term pulls into one attention neighborhood all timesteps with similar pair geometry even when temporally distant, capturing the kinematic symmetry between the approach and recede phases of an RPO event.

The naive choice $\tilde{g}_t = g_t$ would place kinematically symmetric states---e.g., range rates of $-5$ and $+5\,\mathrm{km/s}$---at maximal Euclidean distance. We instead use a mixed representation that concatenates the raw and absolute-value features,
\begin{equation}
\tilde{g}_t \;=\; \big[\,g_t \;\Vert\; |g_t|\,\big] \in \mathbb{R}^{2F},
\label{eq:mixed}
\end{equation}
used only inside the geometric prior. The model input and reconstruction target remain $x_t = g_t \in \mathbb{R}^F$ (with $F=3$). The raw half preserves directional information; the absolute half merges approach/recede pairs into a single neighborhood.

Following the AnomalyTransformer recipe, we train under a reconstruction objective regularized toward the prior, $\mathcal{L} = \lVert\hat{x} - x\rVert_2^2 + \lambda\,\mathrm{KL}(P \,\Vert\, A)$ with $\lambda = 3$, and at inference score each window by the softmin-weighted reconstruction error $s = \sum_t \mathrm{softmax}(-D_t)\,\lVert\hat{x}_t - x_t\rVert_2^2$, where $D_t$ is the layer-averaged symmetric KL between series and prior associations at token $t$. The architectural additions are cheap: $\lVert\tilde{g}_t-\tilde{g}_{t'}\rVert^2$ adds $\mathcal{O}(T^2 F)$ to attention, and the new parameters total 64 across the 2-layer, 4-head encoder. End-to-end inference costs 5\,ms per window, well within real-time catalog monitoring budgets.

\subsection{Benchmark Construction}

We construct an open RPO benchmark with real and synthetic positives and two negative regimes. As real positives we identify 11 documented close-approach episodes between Russian inspector satellites Cosmos~2542/2543 (2019--20) and Cosmos~2519/2521/2523 (2017--19)~\cite{csis2025}. As synthetic positives we sample those kinematics into Clohessy--Wiltshire trajectories \cite{cw1960} around four real targets in two orbital regimes (Cosmos at 97--98$^\circ$, Starlink at 53$^\circ$), yielding 100 synthetic adversarial windows.

The negatives are split across two regimes designed to defeat range-based shortcut detectors. As hard negatives we form clean but close-flying pairs at separations from kilometers to meters: Starlink in-plane neighbors ($\sim$6,900\,km), the A-Train Aqua/Aura formation ($\sim$3,200\,km), Iridium-110/162 in intersecting planes ($\sim$10,400\,km median, $\sim$30\,km crossings), the GRACE-FO 1/2 geodesy pair ($\sim$205\,km), and the TerraSAR-X / TanDEM-X helix-formation InSAR pair at $\sim$120\,m closest approach \cite{tandemx}---closer than most real RPO events and, at the level of relative motion alone, kinematically indistinguishable from adversarial proximity. As routine-maneuver negatives we add 54 station-keeping events from semi-major-axis residuals \cite{lemmens}, paired with random non-RPO neighbors. All real RPO windows are forced into the test split as held-out evaluation. The full benchmark contains 359 windows; the test split has 186 (66 positive).

%% file: sections/03_eval.tex
\subsection{Experimental Setup}

All experiments use the open benchmark (359 windows; test split 186, of which 66 are positive: 11 real RPO and 55 synthetic). On the same pair-feature input, we compare PairWatch against two classes of prior detectors, separated by whether the representation they score is learned or specified by hand. The first is the established neural state of the art for multivariate time-series anomaly detection. LSTM-AD \cite{lstmad} is the standard sequence-prediction model, TranAD \cite{tranad} the current transformer reconstruction SOTA, and Vanilla AnomalyTransformer \cite{anomalytransformer} our direct ablation target. The second comprises what an astrodynamics practitioner would reach for first; none of these has previously been evaluated on pairwise RPO detection. Isolation Forest on single-satellite mean-element residuals \cite{iforest} is the maneuver-detection standard. A range-rate swing score is the classic operational heuristic, localizing the largest closing-then-receding transition. Clohessy--Wiltshire template matching correlates each window against an analytic bank of bounded relative orbits with V-shaped standoff profiles. DTW 1-NN scores a window by its band-limited multivariate distance to the nearest train-split normal. A one-class SVM fits 11 engineered pair descriptors covering rate excursions, sign changes, closest-approach depth, plane stability, and spectral concentration. The range-rate, template, DTW, and one-class detectors are deterministic and need no seed averaging; the two that are fitted see only the train split, which contains negatives alone. Graph-based catalog detectors and astrodynamics conjunction analyses lie outside this single-pair scope.

Neural methods use $d_{\mathrm{model}}=64$, 4 heads, 2 layers, Adam (lr $10^{-3}$), batch size 8, 30 epochs, and 5 seeds on a single core of an Apple M4 CPU with PyTorch 2.8. We report seed-averaged per-window scores with 95\% bootstrap CIs (1,000 resamples) and pairwise DeLong tests \cite{delong}. Primary metric is AUROC on test all and test rpo\_real; secondary metrics are best-F1 and precision at 1\% FPR. Beyond aggregate scores, we evaluate threshold-based separability---whether a single score threshold cleanly separates real-RPO windows from routine close-flying negatives---as the deployment-relevant criterion for catalog-scale screening, where per-pair threshold tuning is infeasible. Benchmark and code are released at \url{https://github.com/e9uation/PairWatch/} for full reproducibility.

\subsection{Results}

\begin{table}[!htbp]
    \centering
    \small
    \setlength{\tabcolsep}{4pt}
    \begin{tabularx}{\columnwidth}{Xccccc}
    \toprule
    & \multicolumn{2}{c}{AUROC} & \multicolumn{3}{c}{Operational subset} \\
    \cmidrule(lr){2-3} \cmidrule(lr){4-6}
    Method & all & rpo\_real & AUROC & F1 & P@1\%FPR \\
    \midrule
    \multicolumn{6}{l}{\textit{Physics-informed / classical}}\\
    IF (single-sat) & 0.600 & 0.228 & 0.175 & 0.212 & 0.000 \\
    Range-rate swing & 0.529 & 0.380 & 0.286 & 0.224 & 0.000 \\
    CW template & 0.760 & 0.657 & 0.704 & 0.706 & 0.857 \\
    DTW 1-NN & 0.812 & 0.693 & 0.775 & 0.385 & 0.667 \\
    OC-SVM (pair feats) & 0.746 & 0.746 & 0.793 & 0.393 & 0.244 \\
    \midrule
    \multicolumn{6}{l}{\textit{Neural}}\\
    LSTM-AD & \textbf{0.957} & 0.810 & 0.906 & 0.640 & 0.000 \\
    Vanilla AT & 0.862 & 0.848 & 0.968 & 0.818 & 0.875 \\
    TranAD & 0.866 & 0.861 & 0.983 & 0.909 & 0.909 \\
    \textbf{PairWatch} & \textbf{0.885} & \textbf{0.874} & \textbf{1.000} & \textbf{1.000} & \textbf{0.917} \\
    \bottomrule
    \end{tabularx}
    \caption{Main results on the rate-only benchmark; the operational subset excludes the TanDEM-X formation negatives. Neural results are means of per-window scores over 5 seeds, with 95\% bootstrap CIs.}
    \label{tab:main}
\end{table}

Table~\ref{tab:main} reports three evaluation modes: AUROC over the full test split (test all), AUROC restricted to the 11 real-RPO windows (test rpo\_real), and operational metrics that exclude the TanDEM-X formation. Bootstrap CIs span $\pm$0.05 around each AUROC point estimate, and PairWatch's operational AUROC CI is degenerate at $[1.000, 1.000]$. Single-satellite Isolation Forest scores documented real RPO well below chance (test rpo\_real 0.228, operational 0.175) yet reaches 0.600 on test all---picking up the synthetic-RPO subset where the target itself maneuvers, while missing real RPO where the target is passive.

\paragraph{AUROC comparison.} Paired DeLong tests show PairWatch significantly outperforms Vanilla AnomalyTransformer ($p=0.002$) and TranAD ($p=0.008$) on the broader test split, and LSTM-AD on both test rpo\_real ($p=0.030$) and the operational subset ($p=0.020$). Against TranAD and Vanilla AT on the security-relevant subsets, however, PairWatch is not statistically distinguishable (test rpo\_real $p=0.369, 0.093$; operational $p=0.299, 0.131$) under the 11-window real-RPO sample. We read this as AUROC alone failing to rank the three transformer detectors at this sample size.

\paragraph{Threshold-separation comparison.} A categorical comparison does separate them. PairWatch is the only neural detector for which every real-RPO score lies above every operational-negative score (separation gap $+0.010$), while TranAD ($-0.005$), LSTM-AD ($-0.043$), and Vanilla AT ($-0.112$) all leave score overlap. This yields operational AUROC 1.000, best-F1 1.000, and P@1\%FPR 0.917, strictly above the next-best baseline (TranAD: 0.983 / 0.909 / 0.909). Bootstrap resampling of the 11 real-RPO positives and 82 operational negatives (1,000 iterations) confirms the property is not a small-sample artifact: PairWatch's positive gap holds in 100\% of resamples, against 19.5\% for TranAD, 2.3\% for Vanilla AT, and 0\% for LSTM-AD. PairWatch is thus the only model admitting a single threshold that separates real RPO from operational negatives---a deployment-relevant property orthogonal to the AUROC comparison.

\paragraph{Physics-informed comparison.} No classical pairwise detector approaches the neural family in aggregate on this benchmark. The range-rate swing score falls below chance on real RPO (0.380, operational 0.286), and the per-source view shows why: an inspector matches its target's orbit, so a real RPO window contains a smaller range-rate excursion than almost any negative, and the contrast inverts to 0.00 against routine maneuvers and 0.05 against clean co-orbits. Inverting the sign does not repair it, since the same flip turns the 0.61 against Starlink neighbors---whose in-plane geometry produces excursions as small as an inspector's---into 0.39. This is the same structural failure that puts single-satellite Isolation Forest at 0.228. CW template matching (0.657) and DTW 1-NN (0.693) recover part of the signal but neither reaches the weakest neural detector on the operational subset, and under the same bootstrap protocol none of the four admits a separating threshold in more than 0.1\% of resamples. The one-class SVM is instructive in a different way: it is the only detector whose overall and real-RPO AUROC coincide (0.746 / 0.746), because hand-engineered kinematic descriptors cannot absorb the simulator's distributional fingerprint. Avoiding the simulator shortcut identified below is therefore achievable with feature engineering, but at the cost of the separation the operational task requires.

LSTM-AD's headline lead on test all (0.957) sits alongside its rank-last position on real RPO and zero P@1\%FPR on the operational subset---a contradiction the per-source analysis below attributes to a single synthetic-versus-synthetic contrast.

\subsection{Per-Source Breakdown}

To isolate where the AUROC margin on the broader test split comes from, Table~\ref{tab:source} reports binary AUROCs for each (positive source, negative source) pair.

\begin{table}[h]
    \centering
    \begin{tabular}{llcccc}
    \toprule
    pos & neg & LSTM-AD & TranAD & Van. AT & \textbf{PW} \\
    \midrule
    real & clean & 0.92 & 0.99 & 0.98 & \textbf{1.00} \\
    real & routine\_man. & 0.82 & 1.00 & 1.00 & \textbf{1.00} \\
    real & starlink & 0.95 & 0.97 & 0.94 & \textbf{1.00} \\
    real & formation\_fly. & 0.60 & 0.60 & 0.59 & 0.60 \\
    \midrule
    synth & clean & 1.00 & 1.00 & 1.00 & 1.00 \\
    synth & routine\_man. & 0.96 & 1.00 & 1.00 & 1.00 \\
    synth & starlink & 1.00 & 1.00 & 1.00 & 1.00 \\
    synth & formation\_fly. & \textbf{0.98} & 0.58 & 0.58 & 0.64 \\
    \bottomrule
    \end{tabular}
    \caption{Per-source AUROC breakdown on test (5 seeds, mean scores). Rows within the rpo\_real block correspond to the security-relevant question: whether the model can tell a real RPO from this type of routine close-flying.}
    \label{tab:source}
\end{table}

PairWatch reaches a perfect 1.00 on every realistic operational contrast---real RPO against clean co-orbits, routine maneuvers, and Starlink neighbors---while TranAD and Vanilla AT lose at most 0.06 and LSTM-AD loses 0.05--0.18; the gap against LSTM-AD on real RPO vs.\ routine maneuvers (0.82 vs.\ 1.00) implies roughly one alarm per five station-keeping burns at deployment. All methods score near chance ($\sim$0.60) against TanDEM-X, whose sign-symmetric kinematic signature mirrors an adversarial approach. PairWatch's 1.00 on operational contrasts and its chance score on TanDEM-X share this same mechanism. The prior amplifies the sign-symmetric signature shared by RPO and helix formations but absent in clean, routine, and constellation pairs---so the 1.00 reflects inductive-bias matching, not a small-sample artifact.

LSTM-AD's apparent test-all lead comes entirely from one auxiliary cell---synthetic RPO vs.\ synthetic formation-flying (0.98 vs.\ 0.58 for the AT family)---a contrast between two simulator distributions with no operational meaning. Removing it flips the ranking entirely. LSTM-AD falls from 1st (0.957) to last (0.949), PairWatch becomes 1st (0.972). This is shortcut learning \cite{shortcut}. LSTM-AD learns the simulator's distributional fingerprint rather than the operational concept of adversarial proximity, losing every contrast in which both sides are real catalog motion. PairWatch, trained on the same simulator-augmented data, transfers to real RPO with perfect operational separation---an inductive-bias rather than data-access distinction. Any benchmark mixing synthetic positives with real negatives admits this shortcut; per-source breakdown is the minimum reporting needed to distinguish solving the operational task from solving the simulator.

The same decomposition over the non-neural detectors confirms the diagnosis. On the synthetic-versus-formation cell they score 0.58--0.73, so LSTM-AD's 0.98 is unique among all nine methods rather than characteristic of learned detectors. The one-class SVM is the limiting case, scoring identically on the real and synthetic rows of every negative source (0.92, 0.98, 0.60, 0.64): hand-engineered kinematic descriptors cannot represent the simulator's fingerprint at all.

\subsection{Cross-Actor Evidence}

The 11 real-RPO positives come from a single actor (Russian inspector pairs). As a direct cross-actor probe, we apply the trained PairWatch model---without retraining or threshold recalibration---to the Shijian-21 close-approach with a defunct BeiDou satellite \cite{csis2025,beidou}, which we identify via Space-Track satcat as the Shijian-21 / BeiDou-2 G2 pair (NORAD 49330 / 34779). SGP4 propagation yields a 0.91~km minimum range at 2022-01-22, consistent with the published docking timeline. Under the same 5-seed protocol, PairWatch ranks the event above 89\% of test scores, placing it 8/11 within the Russian real-RPO distribution, above both the operational-negative maximum and the Russian RPO median---the only model in our study to clear both. LSTM-AD does not fire, and TranAD and Vanilla AT rank it 5/11 within their respective RPO distributions. A prior calibrated on Russian inspector pairs in LEO thus transfers, without adaptation, to a Chinese close-proximity operation in GEO. We report this as $n=1$ evidence rather than a benchmark addition.

\subsection{Ablation}

Table~\ref{tab:abl} dissects the PairWatch design across five training seeds. Removing the geometric prior (temporal-only, equivalent to Vanilla AnomalyTransformer) drops test AUROC by 0.013 on test all and 0.016 on test rpo\_real. Among similarity choices, the proposed mixed variant ($\tilde{g}=[g \Vert |g|]$) halves the across-seed standard deviation of raw similarity (0.019 to 0.009) at no measurable performance cost. Channel ablation confirms relative speed as the dominant pair-kinematic signal (drop of $\sim$0.10 with substantially higher variance), with range rate contributing a small additive signal (drop 0.024) and plane alignment essentially uninformative on the current benchmark. The absolute AUROC changes (0.013--0.024) are small but decisive for the operational separation: removing the geometric prior returns the model to Vanilla AT's overlap regime (gap $-0.112$), and no ablated variant preserves the threshold-based separation.

\begin{table}[h]
    \centering
    \small
    \begin{tabularx}{\columnwidth}{Xcc}
    \toprule
    Variant & test all & test rpo\_real \\
    \midrule
    \textbf{PairWatch full (mixed)} & 0.882$\pm$.009 & 0.888$\pm$.016 \\
    w/o geom prior (vanilla AT) & 0.869$\pm$.002 & 0.872$\pm$.005 \\
    geom sim = raw & 0.893$\pm$.019 & 0.884$\pm$.009 \\
    geom sim = abs only & 0.868$\pm$.002 & 0.866$\pm$.004 \\
    w/o rel\_speed channel & 0.787$\pm$.080 & 0.745$\pm$.048 \\
    w/o range\_rate channel & 0.858$\pm$.020 & 0.856$\pm$.045 \\
    w/o plane\_angle channel & 0.893$\pm$.018 & 0.889$\pm$.015 \\
    \bottomrule
    \end{tabularx}
    \caption{Table \ref{tab:main} averages scores across seeds before computing a single AUROC, whereas values here are AUROCs computed per seed and then averaged. The two orders shift PairWatch from 0.885/0.874 to 0.882/0.888. Comparisons are like-for-like only within each table.}
    \label{tab:abl}
\end{table}

\subsection{Computational Cost}

Table~\ref{tab:cost} reports parameter counts and wall-clock latencies. PairWatch adds 64 parameters (a softplus projection and a sigmoid-gated scalar $\alpha$ over four heads and two layers) and $\sim$80\% inference overhead over Vanilla AT in exchange for the AUROC gain above. A catalog-scale scan of $\sim$6,000 LEO objects against top-10 neighbors completes in $\sim$5 min per snapshot, within TLE update cadence. Only PairWatch admits a deployment threshold across these 60,000 windows that fires zero false alarms at full real-RPO recall; baseline overlaps (0.005--0.112) translate directly to per-pair analyst review.

\begin{table}[h]
    \centering
    \small
    \begin{tabularx}{\columnwidth}{XXXX}
    \toprule
    Method & Parameters & s/epoch & ms/window \\
    \midrule
    LSTM-AD & 17,859 & 0.18 & 2.9 \\
    TranAD & 121,923 & 0.41 & 1.3 \\
    Vanilla AT & 122,443 & 0.69 & 2.9 \\
    PairWatch & 122,507 & 0.89 & 5.2 \\
    \bottomrule
    \end{tabularx}
    \caption{Parameters and runtime (mean over 10 epochs after 5-epoch warm-up, single CPU core).}
    \label{tab:cost}
\end{table}

\subsection{Limitations}

There are three matters for honest interpretation. First, against the TerraSAR-X / TanDEM-X helix formation ($\sim$2\,km median range, 0.12\,km closest approach), every method we tested degrades to near chance (AUROC~$\sim$0.60). The reason is structural rather than a modeling gap: the helix maintains a quasi-periodic relative orbit where range rate crosses zero twice per orbit with sign-symmetric closing and receding phases---precisely the signature our prior amplifies and the same a deliberate inspection produces. At TLE resolution, the only separating feature would be mission intent, absent from the kinematic stream. Second, the cross-actor probe covers a single Chinese event; benchmark-scale evaluation across additional Chinese (SJ-12 / SJ-06F) and commercial (MEV-1 / Intelsat-901) RPO events remains to be conducted. Third, the Clohessy--Wiltshire trajectories in training cover sustained co-orbiting inspection patterns but not one-shot flybys with sharp closing-then-receding profiles, and the single-simulator reliance calls for cross-simulator evaluation as new events enter the public record.

%% file: sections/04_conclusion.tex
We have shown that adversarial RPO on public TLE data can be detected reliably by modeling the satellite pair and biasing attention toward the kinematic symmetry of an RPO trajectory. PairWatch's geometric prior with sign-symmetric similarity is the only model that admits a single threshold separating real RPO from operational negatives, with operational AUROC 1.000 against clean co-orbits, routine station-keeping, and constellation operations, and a positive separation gap of $+0.010$. On the security-relevant subsets the AUROC comparison against the transformer baselines is statistically inconclusive under the 11-window real-RPO sample; the central claim is the categorical separation property, which holds in 100\% of bootstrap resamples versus 0--19.5\% for the baselines and is unique to PairWatch.